\documentclass[preprint,showpacs,preprintnumbers,amsmath,amssymb,prb]{revtex4-1}

\usepackage{amsmath, amsthm, amssymb}
\usepackage{graphicx}
\usepackage{amsmath, amsthm, amssymb}
\usepackage{latexsym}
\usepackage{amssymb}
\usepackage{bm}
\usepackage{dcolumn}
\usepackage{epstopdf}
\usepackage{color}
\usepackage{braket}
\usepackage{mciteplus}

\newcommand{\et}{{\it et\  al.}}

\newcommand{\SHE}{superfluid $^3$He}
\newcommand{\HE}{$^3$He}

\definecolor{cream}{rgb}{1,1,0.7}
\definecolor{orange}{rgb}{1,0.644,0}

\newcommand{\vell}{\mbox{\boldmath$\ell$}}
\newcommand{\vH}{\mbox{\boldmath$H$}}

\def\vH{{\bf H}}

\newcommand{\be}{\begin{equation}}
\newcommand{\ee}{\end{equation}}
\newcommand{\ber}{\begin{eqnarray}}
\newcommand{\eer}{\end{eqnarray}}

\begin{document}
\title{NMR Frequency Shifts and Phase Identification in Superfluid $^3$He}

\author{A. M. Zimmerman$^1$, M. D. Nguyen, and W. P. Halperin}

\affiliation{1:Department of Physics and Astronomy, Northwestern University, Evanston, IL 60208 USA\\ \email{andrewzimmerman2016@u.northwestern.edu}}

\date{\today}

\begin{abstract}
The pressure dependence of the order parameter in \SHE\ is amazingly simple.  In the Ginzburg-Landau regime, {\it i.e.} close to $T_c$,  the square of the order parameter can be accurately measured by its proportionality to NMR frequency shifts and is strictly linear in pressure. This  behavior is  replicated for  \SHE\ imbibed in isotropic and anisotropic silica aerogels.  The proportionality factor is constrained by the symmetry of the  superfluid state and is an important signature of the corresponding superfluid phase.  For the purpose of identifying various new superfluid states in the $p$-wave manifold, the order parameter amplitude of \SHE-A is a useful reference, and this simple pressure dependence greatly facilitates identification. 
\end{abstract}
\maketitle


\section{Introduction}
\begin{figure}[h!]
\hspace*{5 mm}
\includegraphics[width=120mm]{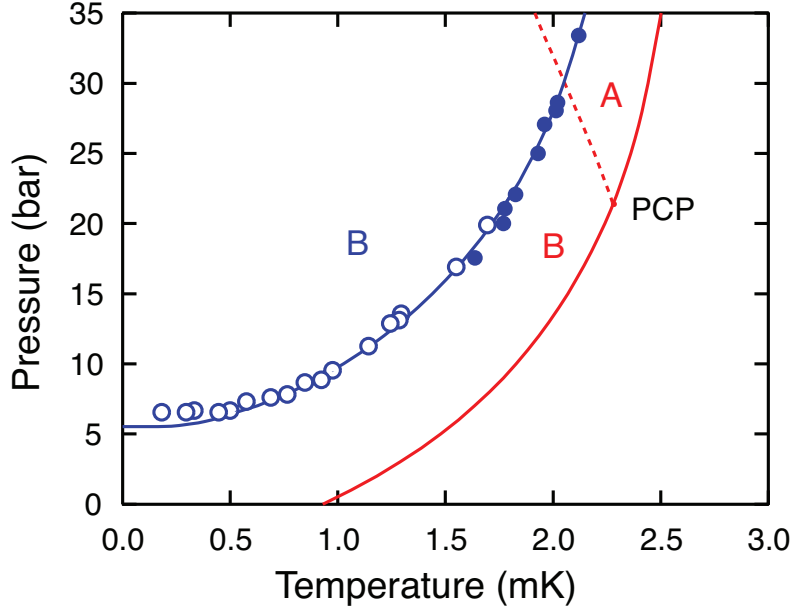}
\caption{\label{fig1} Pressure-temperature phase diagram for \SHE\ in zero magnetic field.  The red traces are for the pure superfluid with  $A$ and $B$ phases and the polycritical point PCP.  The blue curve and data are for  \SHE\ in a 98\% porosity silica aerogel~\cite{Hal.08}}.
\end{figure}

\begin{figure}[t]
\includegraphics[width=120mm]{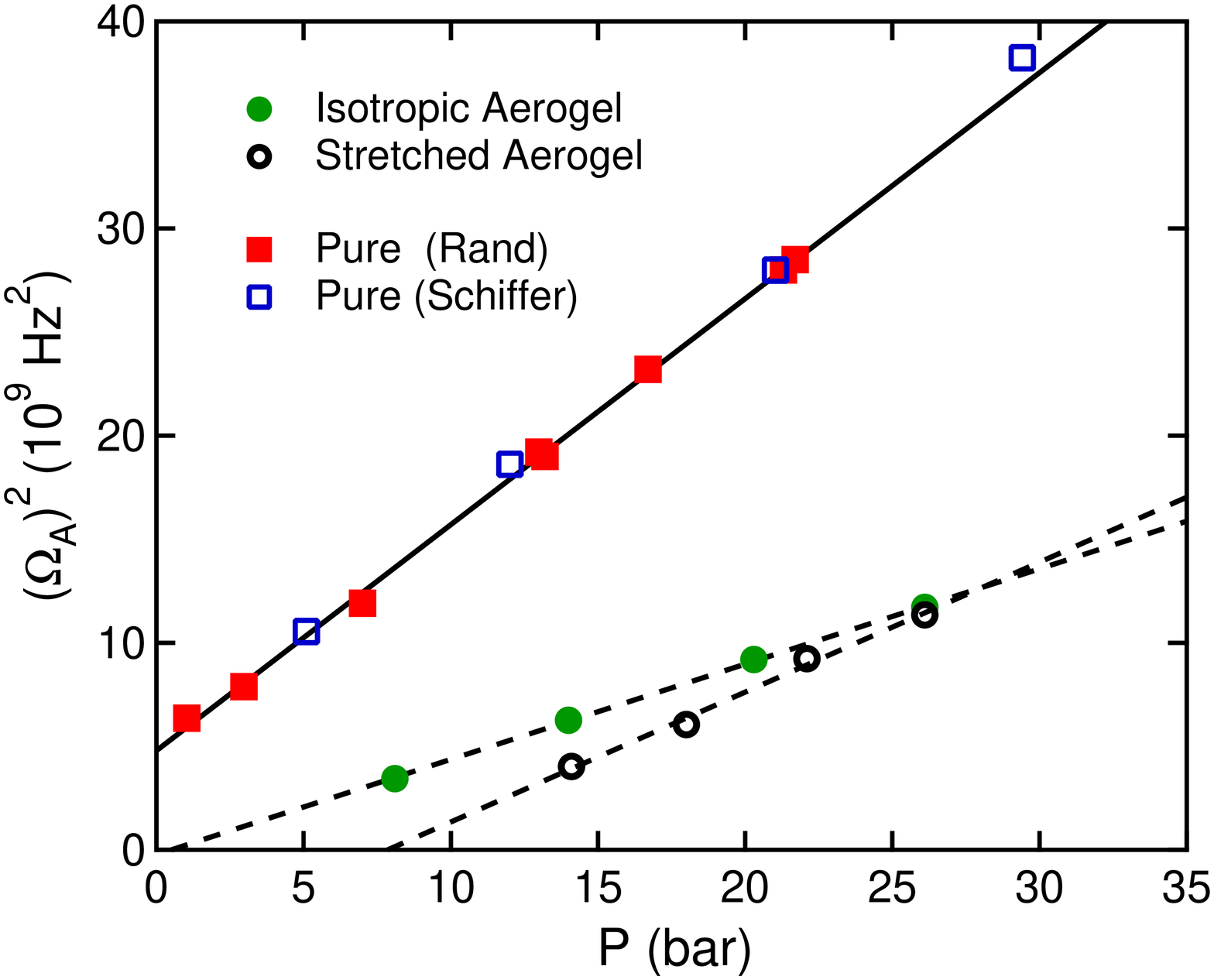}
\caption{\label{fig2} The slope of the reduced temperature dependence of the longitudinal resonance frequency squared near $T_c$ for the $A$ phase in pure \SHE\ taken from Rand~\cite{Ran.96} and Schiffer~\cite{Sch.93,Sch.92};  in 98\% porosity isotropic silica aerogel~\cite{Pol.11}; and in positively strained (stretched) anisotropic silica aerogel~\cite{Pol.12a}. The solid line is a fit to the data from Rand~\cite{Ran.96} and has the functional form $\Omega_A^2(P)/10^{-9}=4.79 \ \text{Hz}^2+(1.09 \ \frac{\text{Hz}^2}{\text{bar}})P$.}
\end{figure}

The pressure dependent \SHE\ transition temperature, $T_c(P)$, Fig.~\ref{fig1}, plays an important role in determining the pressure dependence of most properties of the superfluid, including coherence length, susceptibility, order parameter amplitude,  collective mode frequencies, and the Leggett frequency shift in the transverse nuclear magnetic resonance (NMR) spectrum~\cite{Vol.90}.  However, $T_c$ itself is not directly accessible theoretically and must be treated as an experimental input.    Although $T_c(P)$ is a very non-linear function of pressure, the longitudinal resonance frequency, $\Omega$, measured via the transverse NMR frequency shift, $\Delta\omega(P, T) \propto \Omega^2$, has been found to be simply linear in pressure in the Ginzburg-Landau (GL) regime close to $T_c$. In this limit we measure the slope of $\Omega^2$ versus reduced temperature, Fig.~\ref{fig2}. 
This fact has special significance in that  $\Omega^2$ is proportional to $\Delta^2$, where $\Delta$ is the energy gap that defines the amplitude of the superfluid order parameter as a function of pressure and temperature. In the GL limit, $\Delta^2$ has a well-known linear temperature dependence  $\Delta^2 \propto (1-T/T_c)$.  The linear pressure dependence, however, has no known theoretical basis and has not been discussed previously. A thorough theoretical discussion of the energy gap has been given by Leggett~\cite{Leg.75} and by Vollhardt and W\"olfle~\cite{Vol.90}. 

In the GL limit, the magnitude of the initial slope $\Delta\omega(P)/T_c$  of the temperature dependence of the NMR frequency shift $\Delta\omega(P, T) =  \Delta\omega(P)(1- T/T_c)$  can be measured with high accuracy, provided the NMR magnet homogeneity is of sufficiently high quality. Comparisons between measurements of these frequency shifts in different superfluid phases can give confidence in the identification of the corresponding superfluid states.  This procedure was first exploited by Osheroff~\cite{Osh.74b} on the melting curve to show that the pure superfluid $B$ phase is the isotropic state in comparison with the $A$ phase as the axial state.   A  similar comparison of frequency shifts was used by Rand \et~\cite{Ran.94} to address a proposal by Gould~\cite{Gou.92} that the $A$ phase might  not be a pure axial state. Recently, this method allowed the identification of the superfluid states for \SHE\ in isotropic silica aerogel~\cite{Pol.11} and by Dmitriev \et~\cite{Dmi.15} in anisotropic aerogel. To motivate our discussion in this report, we will review  the relationship between frequency shifts in the known phases of \SHE\  and the symmetry of the superfluid states.

\section{Superfluid Order Parameter}

The order parameter of \SHE\, with maximal amplitude $\Delta$ is a complex second rank tensor, $A_{\mu j}$,  with  spin and orbital  coordinates $\mu$  and $j$ respectively.  In low magnetic field there are two stable phases for pure \SHE, the $A$ and $B$ phases.  A third phase, the Polar ($P$) phase, has only been observed in anisotropic aerogel~\cite{Dmi.15}.  The superfluid phases in aerogel are versions of the $A$, $B$, and $P$ phases for which the order parameter amplitude is reduced, and their stability depends on the properties of the aerogel: specifically, whether the aerogel is uniformly isotropic or anisotropic. For these three phases, the corresponding superfluid $p$-wave order parameters are
\begin{eqnarray}
\begin{aligned}
A^A_{\mu j} & =\Delta_A\hat{d}_{\mu}(\hat{m_1}_j+i\hat{m_2}_j)e^{i\phi}\\
A^B_{\mu j} & =\Delta_BR_{\mu j}e^{i\phi}\\
A^P_{\mu j} & =\Delta_P\hat{d}_{\mu}\hat{p}_je^{i\phi},\\
\end{aligned}
\label{1}
\end{eqnarray}
Where $\hat{d}$ is a vector in spin space; $\hat{m_1}$, $\hat{m_2}$, and $\hat{p}$ are vectors in orbital space; and $R_{\mu j}$ is a relative rotation between the spin and orbital degrees of freedom.

According to Ginzburg-Landau theory, for temperatures just below the second order
thermodynamic transition to the superfluid state, the free energy, $f$, in zero magnetic field can be represented phenomenologically as an expansion in terms of invariants of the order parameter~\cite{Thu.87,Cho.07,Cho.13},
\begin{eqnarray} f &=& -\alpha \mathrm{Tr}(AA^{\dagger})+ g_{z}H_{\mu}(AA^{\dagger})_{\mu\nu}H_{\nu}
+ \beta_{1}|\mathrm{Tr}(AA^{T})|^{2} \nonumber \\ && +\beta_{2}[\mathrm{Tr}(AA^{\dagger})]^{2} 
+\beta_{3}\mathrm{Tr}(AA^{T}(AA^{T})^{*}) \nonumber \\ && +
\beta_{4}\mathrm{Tr}((AA^{\dagger})^{2}) +\beta_{5}\mathrm{Tr}(AA^{\dagger}(AA^{\dagger})^{*}),
\label{GL_free_energy}
\end{eqnarray} 
\noindent
neglecting the dipole energy.  Here, $A^{\dagger}$ and 
$A^{T}$ are, respectively, the Hermitian conjugate and transpose of $A$. There are five fourth-order terms for which the coefficients, $\beta_{i}$,  determine the stable superfluid states.  In the weak-coupling limit we have,
\begin{eqnarray} &\alpha={N(0)\over 3} \left({T\over T_{c}}-1\right),\\ &{\beta_{i}\over\beta_{0}} =
(-1, 2, 2, 2, -2),  i =1, ..., 5,\\ &\beta_{0}={7\zeta (3)\over 240 \pi^{2}} {N(0) \over
(k_{B}T_{c})^{2}}.
\end{eqnarray} 
\noindent
The normal, single-spin density of states at the Fermi energy is  $N(0)$;  $k_{B}$ is the  Boltzmann constant;  and
$\zeta(x)$ is the Riemann zeta function. For the weak-coupling, pure superfluid the isotropic state,
{\it i.e.} the $B$ phase, is  the most stable of all the $p$-wave states.  The existence of the
$A$ phase at high pressure is a consequence  of strong-coupling,  expressed as a pressure dependence of the  $\beta_i$ coefficients deviating from their weak-coupling values, proportional to $T_{c}/T_{F}$ in leading order~\cite{Rai.76}.

The combinations that are relevant here  are those that determine the free energy for a specific superfluid phase, $f_\gamma$, and the corresponding  order parameter amplitudes, {\it i.e.} the maximum energy gaps, $\Delta_\gamma$, where $\gamma$ represents  $A$, $B$, or $P$ phases~\cite{Thu.98}:
\begin{equation}
f_\gamma = k(\alpha/2)\Delta_\gamma^2 = -(\alpha/2)^2/\beta_\gamma,
\label{6}
\end{equation}
\noindent
where $k$ is the dimensionality of the superfluid phase: respectively, 1, 2, and 3, for $P$, $A$, and $B$ phases. Using the Mermin-Stare convention~\cite{Mer.73}, $\beta_{P} \equiv \beta_{12345} \,\,(\equiv \beta_1+\beta_2+\beta_3+\beta_4+\beta_5),\, \,\, \beta_{A} \equiv \beta_{245}$, and $\beta_{B} \equiv \beta_{12} +\beta_{345}/3$.  The condition for the $B$ phase to be stable relative to the $A$ phase is that $\beta_B < \beta_A$.

The energy gaps and heat capacity jumps in the GL limit are related to the $\beta_i$ parameters as:
\begin{equation}
\Delta_{\gamma}^2(T) = \frac{|\alpha(T)|}{k2\beta_{\gamma}};  \,\,\,\,\,\,\,\mathrm{ and}  \,\,\,\,\,\,\, \Delta C_{\gamma} = \frac{\alpha'^2T_c}{2\beta_{\gamma}} ,
\label{7}
\end{equation}
\noindent
where $\Delta C_{\gamma}$ is the heat capacity jump at $T_c$ from the normal state to the corresponding superfluid phase, $\gamma$, and $\alpha'$ is the temperature derivative of $\alpha$.

\section{Phase Identification}
According to Leggett~\cite{Leg.75}, the longitudinal resonance frequencies, $\Omega_{\gamma}$, susceptibilities, $\chi_{\gamma}$  and root mean square order parameter amplitudes, $\Psi_{\gamma}$,  can be compared for $A$ and $B$ phases, which we extend  to a comparison of the $A$ and $P$ phases as follows:
\begin{equation}
\frac{5}{2}=\frac{\Omega_{B}^{2}}{\Omega_{A}^{2}}\frac{\chi_{B}}{\chi_{A}}\frac{\Psi_{A}^{2}}{\Psi_{B}^{2}};\,\,\,\,\,\,\,\,\,\,\,
2=\frac{\Omega_{P}^{2}}{\Omega_{A}^{2}}\frac{\Psi_{A}^{2}}{\Psi_{P}^{2}}.
\label{8}
\end{equation}
\noindent
The ratios of $\Psi^2$, can be replaced by the ratios of the square of the maximum energy gaps, $\Delta^2$.
\begin{equation}
\frac{\Psi_{A}^{2}}{\Psi_{B}^{2}} = \frac{2}{3}\frac{\Delta_{A}^{2}}{\Delta_{B}^{2}} = \frac{\beta_B}{\beta_A} = \frac{\Delta C_A}{\Delta C_B} ;\,\,\,\,\,\,\,\,\,\frac{\Psi_{A}^{2}}{\Psi_{P}^{2}} =2\frac{\Delta_{A}^{2}}{\Delta_{P}^{2}} = \frac{\beta_P}{\beta_A} = \frac{\Delta C_A}{\Delta C_P}.
\label{9}
\end{equation}
\noindent
From Eq.~\ref{8} and~\ref{9} this leads to,
\begin{equation}
\frac{\Omega_{B}^{2}}{\Omega_{A}^{2}}=\frac{5}{2}\frac{\chi_{A}}{\chi_{B}}\frac{\beta_{A}}{\beta_{B}}=\frac{5}{2}\frac{\chi_{A}}{\chi_{B}}\frac{\Delta C_{B}}{\Delta C_{A}};\,\,\,\,\,\,\,\,\,\,\,
\frac{\Omega_{P}^{2}}{\Omega_{A}^{2}}=2\frac{\beta_{A}}{\beta_{P}}=2\frac{\Delta C_{P}}{\Delta C_{A}}.
\label{10}
\end{equation}

The relationships given in Eq.~\ref{8} and Eq.~\ref{10} can be used for phase identification in different regimes. At high pressure, near the PCP, we can take $\beta_{B}=\beta_{A}$ for pure \SHE . This results in $\Omega_B^2/\Omega_A^2 = (5/2)\chi_A/\chi_B$, which was used in pure \SHE\ ~\cite{Osh.74b, Ran.94} and was shown to hold for isotropic aerogel~\cite{Pol.11}. At low pressure, in the weak coupling limit,  $\beta_B/\beta_A = 5/6$ and $\beta_P/\beta_A = 3/2$, resulting in  $\Omega_B^2/\Omega_A^2 = 3\chi_A/\chi_B$ and $\Omega_P^2/\Omega_A^2 = 4/3$ as used in anisotropic aerogel~\cite{Dmi.15}. Outside of these limits, more details of the relevant $\beta$ parameters are needed.
\section{Transverse NMR}
It is illuminating to express these relationships directly in terms of the transverse NMR frequency shifts. With transverse NMR we can identify the frequency shifts for energetically stable textures with correspondingly minimum  or  maximum dipole energy that we call dipole-locked or dipole-unlocked and are specified by choice of  $\theta = (\hat{\vell},\hat{\vH})$ and NMR tip angle $\beta$, where $\hat{\vell}$ is the quantization axis of angular momentum.  These frequency shifts are expressed in terms of form factors, $F_{\gamma}(\theta,\beta)$,  and the Larmor frequency, $\omega_L$,
\begin{equation}
\omega_{\gamma}^2 = \omega^2_{L} + F_{\gamma}(\theta,\beta)\Omega^2_{\gamma}. 
\label{11}
\end{equation}
\noindent
At magnetic fields greater than the dipole field ($\approx 3$\, mT), we can write Eq.~\ref{11} as,
 \begin{equation}
 \Delta\omega_{\gamma} = F_{\gamma}\Omega_{\gamma}^2/2\omega_L.
 \label{12}
 \end{equation}
For the $A$ phase ~\cite{Gon.80, Yur.93b} (dipole locked, $\theta = \pi/2$),
 \begin{equation}
\Delta\omega_A(\theta,\beta)=\frac{{\Omega_A}^2}{2\omega_L} \left(-\cos\beta + \left(\frac{7}{4}\cos\beta+\frac{1}{4}\right)  \sin^2 \theta \right).
\label{13}    
\end{equation} 
For the $B$ phase~\cite{Bri.75, Cor.78, Dmi.99} (dipole locked),
\begin{eqnarray}
 \Delta\omega_B(0,\beta)&\approx&0,\hspace{35pt}\beta < 104^{\circ}\label{1}\\
\Delta\omega_B(0,\beta)&=&-\frac{8}{15}\frac{\Omega_{B}^{2}}{2\omega_{L}}(1+4\cos\beta).\hspace{16pt}\beta > 104^{\circ}, \label{14+15}
 \end{eqnarray}
\noindent
and for (dipole unlocked),
\begin{eqnarray}
\Delta\omega_B(\frac{\pi}{2},\beta)&=&\frac{\Omega_{B}^{2}}{2\omega_{L}}(\cos\beta-\frac{1}{5}),\hspace{20pt}\beta < 90^{\circ}\label{16}\nonumber\\
\Delta\omega_B(\frac{\pi}{2},\beta)&=&-\frac{1}{5}\frac{\Omega_{B}^{2}}{2\omega_{L}}(1+\cos\beta).\hspace{20pt}\beta > 90^{\circ} 
\label{16}
\end{eqnarray}
For the polar phase~\cite{Dmi.15,Wim.15} (dipole locked, $\theta = \pi/2$),
\begin{equation}
\Delta\omega_P(\theta,\beta) = \frac{{\Omega_P}^2}{2\omega_L} \left(\cos\beta  - \frac{\cos^2\theta}{4}(5\cos\beta-1)\right).
\label{17}    
\end{equation} 
It is most convenient to measure frequency shifts in the configurations: $\theta=\pi/2$ and $\beta=0$, for which $F_A=1$ (locked), $F_B = 4/5$ (unlocked), and  $F_P = 1$ (locked). In these cases, from Eq.~\ref{10}, \ref{11}, we have:
\begin{eqnarray}
\frac{\Delta\omega_B}{\Delta\omega_A}&=&2\frac{\beta_{A}}{\beta_{B}}\frac{\chi_{A}}{\chi_{B}}=2\frac{\Delta C_{B}}{\Delta C_{A}}\frac{\chi_{A}}{\chi_{B}}\nonumber\\
\frac{\Delta\omega_P}{\Delta\omega_A}&=&2\frac{\beta_{A}}{\beta_{P}}=2\frac{\Delta C_{P}}{\Delta C_{A}}.
\label{18}
\end{eqnarray}
The measured NMR frequency shifts relative to the $A$ phase are twice the ratio of the corresponding heat capacity jumps, including strong coupling corrections. This fact could be useful since the frequency shifts are more easily and more precisely measurable than the heat capacity.  

\section{Conclusions}

The identification of $A$ and $B$ phases is well-established for pure \HE\ and does not need to be revisited.  However,  the above expressions are important for recent attempts to identify the phases of \SHE\ in confinement in pores, slabs, and in aerogels, especially for various forms of polar distorted $A$ and $B$ phases, and the $P$ phase. The principal complication in using this approach is the accessibility of the $\beta$ parameters for the corresponding phases.  Measurements of the heat capacity would be ideal, but are rather involved, and may not be as precise for \SHE\ in aerogel~\cite{Cho.04}  as for pure \HE. 

Another approach is to scale the appropriate $\beta$ parameters from their pure \HE\ values~\cite{Cho.07,Wim.18} using theoretical models such as that of Thuneberg \et~\cite{Thu.98}, a method that was used by Pollanen \et~\cite{Pol.11} with isotropic silica aerogel.  Recently, the Thuneberg \et\,\,theory was been extended beyond the GL limit by Wiman and Sauls~\cite{Wim.18},  including  temperature dependence for the $\beta$ parameters.   Application of the theoretical models to obtain the $\beta$'s requires an experimental determination of the appropriate model parameters, such as the quasiparticle mean free path, $\lambda$, and the correlation length, $\xi_a$.  These can be determined from the experimental pressure-temperature phase diagram, for which the theory gives an accurate description  (shown in Fig.~\ref{fig1} by the solid blue curve).  However, the pressure dependence of $T_c$  at low pressure is particularly important for extracting these parameters, since in this limit we know that their strong coupling contributions go to zero as $T \rightarrow 0$~\cite{Rai.76}.  We propose that this region can be explored using the linear pressure dependence of $\Omega^2$. By extrapolating to the pressure $P_0$, where $\Omega^2(P_0)=0$, we find a critical pressure, at which $T_c(P_0)=0$, constraining the low temperature phase diagram. In Fig.~\ref{fig2} we show that this behavior is a general feature  of \SHE\ in both its pure and confined forms.  This procedure was described and implemented for  chiral phases in positively strained silica aerogel~\cite{Pol.12a}.

Finally, it has been predicted~\cite{Wim.15,Wim.18}, and there is experimental evidence~\cite{Dmi.15}, that there are transitions between ESP phases as a function of pressure. In order to identify the existence of these phases and transitions between them,  it is useful to examine the pressure dependence of the order parameter through measurement of the NMR frequency shift.  An abrupt deviation from the linear behavior shown in Fig.~\ref{fig2} is the signal for such a transition.   

\begin{acknowledgements}
The authors acknowledges contributions from Jim Sauls, Johannes Pollanen, Jia Li, and  Josh Wiman. Research was supported by the National Science Foundation, Division of Materials Research: DMR-1602542.
\end{acknowledgements}

\end{document}